\documentclass[letter,twocolumn]{jpsj3}
\usepackage{graphicx}
\usepackage{txfonts}
\usepackage{bm}% bold math
\usepackage{amsmath}
\usepackage{color}

\title{Crystalline Electronic Field in Rare-Earth Based Quasicrystal and Approximant: 
Analysis of Quantum Critical Au-Al-Yb Quasicrystal and Approximant}

\author{Shinji Watanabe and Mina Kawamoto}
\inst{Department of Basic Sciences, Kyushu Institute of Technology, Kitakyushu, Fukuoka 804-8550, Japan
}
\abst{
On the basis of the point charge model, we formulate the crystalline electronic field (CEF) Hamiltonian $H_{\rm CEF}$ in the rare-earth based quasicrystal (QC) and approximant crystal (AC) with ligand ions located at pseudo 5-fold configurations by using the operator equivalent method. 
By setting the total angular momentum $J=7/2$, the CEF in the quantum critical QC Au$_{51}$Al$_{34}$Yb$_{15}$ and the 1/1 AC Au$_{51}$Al$_{35}$Yb$_{14}$ is analyzed with consideration for the effect of Al/Au mixed sites. 
We find that the ratio of the valences of ligand ions $x=Z_{\rm Al}/Z_{\rm Au}$ plays an important role 
{in}
 characterizing the CEF ground state. 
As $x$ decreases from $x=3$, the 4f wave function of the CEF ground state with the flat shape lying in the mirror plane is deformed around $x\approx 0.8$ to the flat shape   perpendicular to the pseudo 5-fold axis at $x=0$. 
The formulated $H_{\rm CEF}$ by $J$ is generally applicable to rare-earth-based QCs and ACs, which is useful to analyze the CEF.
}

\begin{document}

\def\gsim{\mathop {\vtop {\ialign {##\crcr 
$\hfil \displaystyle {>}\hfil $\crcr \noalign {\kern1pt \nointerlineskip } 
$\,\sim$ \crcr \noalign {\kern1pt}}}}\limits}
\def\lsim{\mathop {\vtop {\ialign {##\crcr 
$\hfil \displaystyle {<}\hfil $\crcr \noalign {\kern1pt \nointerlineskip } 
$\,\,\sim$ \crcr \noalign {\kern1pt}}}}\limits}

\maketitle

%\section{Introduction}

Discovery of the rare-earth based quasicrystal (QC) has opened a new research field of strongly-correlated electrons on the QC.   
The QC Au$_{51}$Al$_{34}$Yb$_{15}$ exhibits unconventional quantum critical phenomena at ambient pressure, which surprisingly persists even under pressure at least up to $P=1.6$~GPa~\cite{Deguchi,Watanuki}. 
The quantum criticality in physical quantities such as the magnetic susceptibility, the specific-heat coefficient, the resistivity, and the NMR relaxation rate and its robustness against pressure have been shown to be explained by the theory of critical Yb-valence fluctuations~\cite{WM2010,WM2013,WM2016,WM2018,WM2020}. 
Recently, a sharp change in the Yb valence has been observed in the QC (Au$_{1-y}$Cu$_{y}$)$_{51}$ (Al$_{1-x}$Ga$_{x}$)$_{34}$Yb$_{15}$ at $x=y=0$, which gives the direct evidence of the quantum valence criticality~\cite{Imura}.

\begin{figure}[t]
\includegraphics[width=7cm]{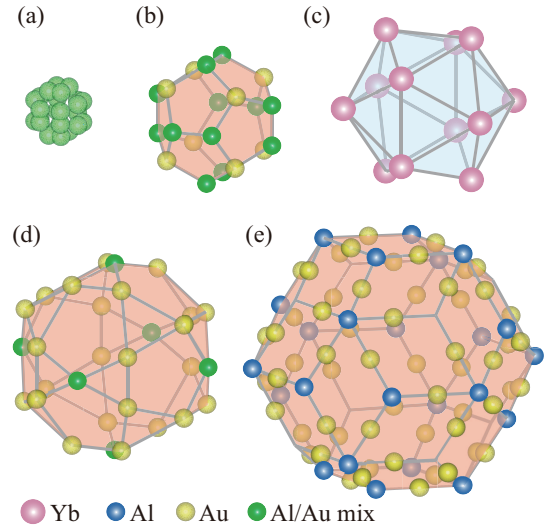}
\caption{(Color online) 
Tsai-type cluster consists of (a) the 
%%%%%
%\textcolor{red}
{cluster center},
%%%%%
 (b) the 
%%%%%
%\textcolor{red}
{first}
%%%%%
 shell, (c) the 
%%%%%
%\textcolor{red}
{second}
%%%%% 
  shell, (d) the 
%%%%%
%\textcolor{red}
{third}
%%%%%  
   shell, and (e) the 
%%%%%
%\textcolor{red}
{fourth}
%%%%%   
    shell 
with Yb (pink), Al (blue), Au (yellow), and Al/Au mixed site (green).
} 
\label{fig:YbAlAu}
\end{figure}

The QC Au$_{51}$Al$_{34}$Yb$_{15}$ consists of the Tsai-type cluster which has concentric shell structures shown in Fig.~\ref{fig:YbAlAu}. In Fig.~\ref{fig:YbAlAu}(c), Yb atoms are located at each vertex of the icosahedron, forming the Yb 12 cluster. 
In the 
%%%%%
%\textcolor{red}
{cluster center}
%%%%%
 [Fig.\ref{fig:YbAlAu}(a)], the 
%%%%%
%\textcolor{red}
{1st}
%%%%%
  shell [Fig.\ref{fig:YbAlAu}(b)], and the 
%%%%%
%\textcolor{red}
{3rd}
%%%%%   
  shell [Fig.\ref{fig:YbAlAu}(d)], 
Al/Au mixed sites exist with existence ratio 7.8\%/8.9\%, 62\%/38\%, and 59\%/41\%, respectively~\cite{Ishimasa}. 
There also exists the 1/1 approximant crystal (AC) Au$_{51}$Al$_{35}$Yb$_{14}$ with periodic arrangement of the unit cell where the Tsai-type cluster is located at the corner and center in the body-centered cubic (bcc) lattice. 
%The lattice constant is $a=14.5$~\AA. 

%%%%%%%%%%%%  Fig.1  %%%%%%%%%%%%%%%%%%%%%%%%%%%%%%
%%%%%%%%%%%%%%%%%%%%%%%%%%%%%%%%%%%%%%%%%%%%%%%%

\begin{figure}[t]
\includegraphics[width=7cm]{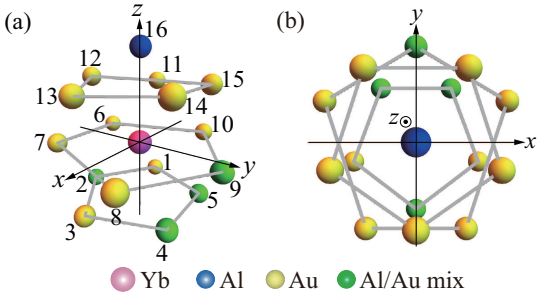}
\caption{(Color online) (a) Local configuration around the Yb atom (pink). The number labels surrounding Al (blue), Au (yellow), and Al/Au mixed site (green). 
(b) Top view from the $z$ direction of (a).  
} 
\label{fig:Yb_local}
\end{figure}

Although understanding of the origin of the quantum criticality has proceeded, the detail of the crystalline electronic field (CEF) in the QC and AC has been unresolved.   
Recently, the CEF in rare-earth based ACs with the bcc lattice composed of the Tsai-type cluster has been studied experimentally. 
The specific-heat and magnetic-susceptibility measurements in Zn$_{85.5}$Sc$_{11}$Tm$_{3.5}$~\cite{Jazbec} and the neutron measurement in Cd$_6$Tb~\cite{Das} have reported that the CEF ground state is $|J=6, J_z=\pm 6\rangle$.
Here, $J$ is the total angular momentum and both Tm$^{3+}$ (4f$^{12}$) and Tb$^{3+}$ (4f$^8$) have the $J=6$ ground state according to the Hund's rule. 
The wave function $\Phi_{\pm}(\hat{\bm r})=\langle\hat{\bm r}|J=6, J_z=\pm 6\rangle$ is lying in the plane perpendicular to the pseudo 5-fold axis (see the $xy$ plane and $z$ axis in Fig.~\ref{fig:Yb_local}, respectively). 

Very recently, inelastic neutron measurement combined with the analysis of the CEF by the point-charge model in the AC Au$_{70}$Si$_{17}$Tb$_{13}$ has suggested that the wave function of the CEF ground state is lying in the 
%mirror plane (i.e., the $yz$ plane in Fig.~\ref{fig:Yb_local})
%%%%%
%\textcolor{red}
{plane parallel to the pseudo 5-fold axis (i.e., the $z$ axis in Fig.~\ref{fig:Yb_local})}~\cite{Hiroto}.
%%%%%
 This is in sharp contrast to that in ref.~\citen{Das} albeit both materials have the Tb-based bcc lattice composed of the Tsai-type cluster. 
It is also noted that magnetic long-range orders have been observed in the AC 
Cd$_6$R (R=Nd, Sm, Gd, Tb, Dy, Ho, Er, and Tm)~\cite{Tamura2010,Mori,Tamura2012} and Au-SM-R (SM=Si, Ge, and Sn; R=Gd, Tb, Dy, and Ho)~\cite{Hiroto2013,Hiroto2014} which consist of the Tsai-type cluster. 
The importance of the magnetic anisotropy in the magnetic orders has been reported~\cite{Tamura2010,Sugimoto,Ishikawa,Sato2019}. 
To clarify its origin, theoretical analysis of the CEF is highly desired. 
These circumstances motivate us to study theoretically the CEF in the rare-earth based QC and AC.  

In this letter,  
we formulate the CEF Hamiltonian $H_{\rm CEF}$ in the rare-earth based QC and AC by the total angular momentum $J$ using the operator equivalent method. 
By setting $J=7/2$, we analyze the CEF of the QC Au$_{51}$Al$_{34}$Yb$_{15}$ and the AC  
taking into account the effect of the Al/Au mixed site.  
We clarify the CEF states and find that the ratio of valences of  
the screened ligand ions plays the important role 
%%%%%
%\textcolor{red}
{in}
%%%%%
characterizing the CEF ground state. 
%Since $H_{\rm CEF}$ is formulated for general $J$, our formulation is applicable to the other rare-earth based QC and AC with the Tsai-type cluster, which is expected to contribute to the understanding of the CEF in these systems. 
%we further develop  the analysis of the CEF in the QC Yb$_{15}$Al$_{34}$Au$_{51}$ which was qualitatively discussed in Ref.~\citen{WM2018}. 

%\section{Formulation of the CEF Hamiltonian by equivalent operator method}

Let us start with the formulation of $H_{\rm CEF}$. 
As a first step of analysis, we study the CEF on the basis of the point charge model. 
The CEF Hamiltonian for the Yb$^{3+}$ ion is given by $H_{\rm CEF}=|e|V_{\rm cry}({\bm r})$ in the hole picture. 
Since Yb$^{3+}$ has 4f$^{13}$ configuration, this state is regarded to have a hole in the closed shell with 4f$^{14}$ configuration for Yb$^{2+}$. Hence, the charge of the 4f hole at Yb$^{3+}$ is set to be $+|e|$. 
The potential $V_{\rm cry}({\bm r})$ is given by 
\begin{eqnarray}
V_{\rm cry}({\bm r})&=&\sum_{i=1}^{16}\frac{q_i}{\left|{\bm R}_i-{\bm r}\right|},
\nonumber
\\
&=&\sum_{i=1}^{16}\sum_{\ell=0}^{\infty}\sum_{m=-\ell}^{\ell}\frac{q_i}{R_i}\left(\frac{r}{R_i}\right)^{\ell}
\frac{4\pi(-1)^m}{2l+1}Y^{m}_{\ell}(\theta_i, \varphi_i)Y^{-m}_{\ell}(\theta, \varphi),  
\label{eq:Vcry}
\end{eqnarray}
where $Y^{m}_{\ell}(\theta,\varphi)$ is the spherical harmonics with the azimuthal quantum number $\ell$ and the magnetic quantum number $m$. 
Here, ${\bm R}_i=(R_i, \theta_i, \phi_i)$ is the position vector of the ligand ions of Al$^{Z_{\rm Al}+}$ and Au$^{Z_{\rm Au}+}$ 
with $Z_{\rm Al}$ and $Z_{\rm Au}$ being valences of Al and Au respectively 
for $i=1$-$16$ shown in Fig.~\ref{fig:Yb_local}(a), which surround the Yb ion in the Tsai-type cluster in the QC and AC~\cite{Matsukawa2014,Ishimasa}. 
Since the existence ratio of the 
%%%%%
%\textcolor{red}
{cluster center}
%%%%%
 is quite small, here we neglect the 
%%%%%
%\textcolor{red}
{cluster center}
%%%%%
  [Fig.~\ref{fig:YbAlAu}(a)]. 
Then $q_i$ is given by $q_i=Z_{\rm Al}|e|$ and $q_i=Z_{\rm Au}|e|$ for each site in Fig.~\ref{fig:Yb_local}(a). 

%%%%%%%%%%%%  Fig.2  %%%%%%%%%%%%%%%%%%%%%%%%%%%%%%
%%%%%%%%%%%%%%%%%%%%%%%%%%%%%%%%%%%%%%%%%%%%%%%%

By using the Wigner-Eckart theorem~\cite{Wigner}, $H_{\rm CEF}$ can be expressed by Stevens operators~\cite{Stevens} as 
\begin{eqnarray}
H_{\rm CEF}&=&B_{2}^{0}O_{2}^{0}
+B_{2}^{2}O_{2}^{2}+B_{2}^{2}(s)O_{2}^{2}(s)
+B_{2}^{1}O_{2}^{1}+B_{2}^{1}(s)O_{2}^{1}(s)
\nonumber
\\
&+&B_{4}^{0}O_{4}^{0}
+B_{4}^{4}O_{4}^{4}+B_{4}^{4}(s)O_{4}^{4}(s)
+B_{4}^{3}O_{4}^{3}+B_{4}^{3}(s)O_{4}^{3}(s)
\nonumber
\\
&+&B_{4}^{2}O_{4}^{2}+B_{4}^{2}(s)O_{4}^{2}(s)
+B_{4}^{1}O_{4}^{1}+B_{4}^{1}(s)O_{4}^{1}(s)
\nonumber
\\
&+&B_{6}^{0}O_{6}^{0}
+B_{6}^{6}O_{6}^{6}+B_{6}^{6}(s)O_{6}^{6}(s)
+B_{6}^{5}O_{6}^{5}+B_{6}^{5}(s)O_{6}^{5}(s)
\nonumber
\\
&+&B_{6}^{4}O_{6}^{4}+B_{6}^{4}(s)O_{6}^{4}(s)
+B_{6}^{3}O_{6}^{3}+B_{6}^{3}(s)O_{6}^{3}(s)
\nonumber
\\
&+&B_{6}^{2}O_{6}^{2}+B_{6}^{2}(s)O_{6}^{2}(s)
+B_{6}^{1}O_{6}^{1}+B_{6}^{1}(s)O_{6}^{1}(s),
\label{eq:Vcry2}
\end{eqnarray}
where $B_{\ell}^{m}$ is defined as 
$B_{2}^{m}=-|e|C_{2}^{m}\langle r^{2}\rangle\alpha_Jh_2^{m}$, 
$B_{4}^{m}=-|e|C_{4}^{m}\langle r^{4}\rangle\beta_Jh_4^{m}$, and 
$B_{6}^{m}=-|e|C_{6}^{m}\langle r^{6}\rangle\gamma_Jh_6^{m}$. 
Here, $C_{\ell}^{m}$ is given by 
$C_2^0=\sqrt{\pi/5}$, $C_2^1=2\sqrt{6\pi/5}$, $C_2^2=\sqrt{6\pi/5}$, 
$C_4^0=\sqrt{\pi}/12$, $C_4^1=\sqrt{5\pi}/3$, $C_4^2=\sqrt{10\pi}/6$, $C_4^3=\sqrt{140\pi}/6$, $C_4^4=\sqrt{70\pi}/12$,  
$C_6^0=\sqrt{13\pi}/104$, $C_6^1=\sqrt{546\pi}/52$, $C_6^2=4\sqrt{5460\pi}/793$, $C_6^3=\sqrt{5460\pi}/104$, $C_6^4=3\sqrt{182\pi}/104$, $C_6^5=\sqrt{9009\pi}/52$, and $C_6^6=\sqrt{12012\pi}/208$.               
The Stevens factors $\alpha_J$, $\beta_J$, and $\gamma_J$ are given by $\alpha_J=2/63$, $\beta_J=-2/1155$, and $\gamma_J=4/27027$  for Yb$^{3+}$~\cite{Hutchings}. 
The expectation value by the radial part of the wave function is given by 
$\langle r^2\rangle=0.1826~\AA^2$, $\langle r^4\rangle=0.0854~\AA^4$,  and $\langle r^6\rangle=0.0863~\AA^6$ on the basis of the Dirac-Fock calculation for Yb$^{3+}$~\cite{Freeman}. 
Here, $h_{\ell}^{m}$ is given by
\begin{eqnarray}
h_{\ell}^{m}&=&{\rm Re}\left[\sum_{i=1}^{16}\frac{q_i}{R_i^{\ell+1}}(-1)^{m}Y_{\ell}^{m}(\theta_i,\varphi_i)\right]. 
\label{eq:hlm}
\end{eqnarray}
In Eq.~(\ref{eq:Vcry2}), 
$B_{\ell}^{m}(s)$ is obtained by inputting $h_{\ell}^{m}(s)$ defined by 
\begin{eqnarray}
h_{\ell}^{m}(s)&=&{\rm Im}\left[\sum_{i=1}^{16}\frac{q_i}{R_i^{\ell+1}}(-1)^{m}Y_{\ell}^{m}(\theta_i,\varphi_i)\right],
\end{eqnarray}
instead of $h_{\ell}^{m}$ into each $B_{\ell}^{m}$ above. 

Stevens operators $O_{\ell}^{m}$ and $O_{\ell}^{m}(s)$ are expressed by operators of the total angular momentum $J_z$, $J_{+}\equiv (J_x+J_y)/2$, and $J_{-}\equiv (J_x-J_y)/(2i)$  (see supplemental materials)~\cite{SM,Stevens,Hutchings,rotter}.  
The Hund's rule tells us that the ground state of Yb$^{3+}$ is given by $J=7/2$. 
%%%%%
%\textcolor{red}
%{
We see that all the terms in the right hand side (r.h.s.) of Eq. (\ref{eq:Vcry2}) remain. This is in sharp contrast to the CEF Hamiltonian in the heavy-fermion systems specified by the crystal point group, where most terms are canceled owing to the high symmetry~\cite{WM2018,Hiroto}. This is characteristic of $H_{\rm CEF}$ in the QC and AC. 
%}
%%%%%

By diagonalizing the $8\times 8$ CEF Hamiltonian $H_{\rm CEF}$ with the basis $|J_z\rangle$ with $J_z=7/2, 5/2, 3/2, 1/2, -1/2, -3/2, -5/2$ and $-7/2$ for $J=7/2$, we obtain the CEF energies and the eigenvectors. 
The four sets of eigenstates with Kramers degeneracy are obtained as 
\begin{eqnarray}
|\psi_{n+}\rangle&=&a_{n1}\left|\frac{7}{2}\right\rangle+a_{n2}\left|\frac{5}{2}\right\rangle+a_{n3}\left|\frac{3}{2}\right\rangle+a_{n4}\left|\frac{1}{2}\right\rangle
\nonumber
\\
&+&a_{n5}\left|-\frac{1}{2}\right\rangle+a_{n6}\left|-\frac{3}{2}\right\rangle+a_{n7}\left|-\frac{5}{2}\right\rangle+a_{n8}\left|-\frac{7}{2}\right\rangle,
\label{eq:WFup}
\\
|\psi_{n-}\rangle&=&a_{n8}^{*}\left|\frac{7}{2}\right\rangle-a_{n7}^{*}\left|\frac{5}{2}\right\rangle+a_{n6}^{*}\left|\frac{3}{2}\right\rangle-a_{n5}^{*}\left|\frac{1}{2}\right\rangle
\nonumber
\\
&+&a_{n4}^{*}\left|-\frac{1}{2}\right\rangle-a_{n3}^{*}\left|-\frac{3}{2}\right\rangle+a_{n2}^{*}\left|-\frac{5}{2}\right\rangle-a_{n1}^{*}\left|-\frac{7}{2}\right\rangle, 
\label{eq:WFdw}
\end{eqnarray}
where $a_{ni}$ is the complex numbers satisfying $\sum_{i=1}^{8}|a_{ni}|^2=1$ for $n=0,1,2$, and 3~\cite{WM2018}. 

%\section{Results}

%\subsection{The case of Al 100\%}

First we discuss the case that Al is 100\% occupied at the Al/Au mixed sites as a representative case, which was shown to have the highest probability among possible Al/Au configurations [see Fig.~\ref{fig:AlAu}(b)]~\cite{WM2018}. 
The effect of the Al/Au mixed sites will be discussed later.

In metallic crystals, the valences of Al and Au are known to be $Z_{\rm Al}=3.0$ and $Z_{\rm Au}=1.0$ respectively 
%in literatures~\cite{Pearson,Mizutani}.
in normal metal~\cite{Pearson} and alloyed AC~\cite{Mizutani}. 
In reality, screening effect by conduction electrons is expected to reduce $Z_{\rm Al}$ and $Z_{\rm Au}$ from the above values~\cite{Hiroto}. 
Hence, we analyze the CEF for $Z_{\rm Al}=xZ_{\rm Au}$ by changing $x$ for $0 \le x\le 3$. 
Here we set $Z_{\rm 
%%%%%
%\textcolor{red}
{Au}
%%%%%
}=0.01776159$ so as the CEF excitation energy $\Delta\equiv E_1-E_0$ at $x=3$ to be $20$~K 
estimated from the entropy obtained by 
%integrating the electric specific heat by temperature $S(T)=\int_0^TdT'C_{\rm e}(T')/T'$ 
in the QC Au$_{51}$Al$_{34}$Yb$_{15}$~\cite{Deguchi,D_PC}. 
%%%%%
%\textcolor{red}
{So far, direct experimental verification of these choices of $x$ and $Z_{\rm Au}$ has not been reported. However, if one compares the following Fig.~\ref{fig:E_x} with the data of the future neutron measurement, it is possible to identify $x$ and $Z_{\rm Au}$ experimentally, which will be discussed in detail later.}
%%%%%

%%%%%%%%%%%%  Fig.3  %%%%%%%%%%%%%%%%%%%%%%%%%%%%%%
\begin{figure}[t]
\includegraphics[width=7cm]{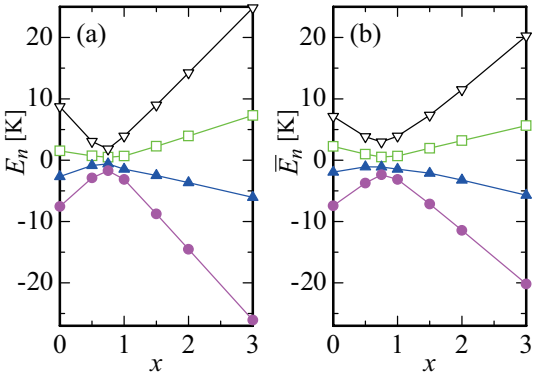}
\caption{(Color online) 
(a) The $x$ dependence of the CEF energy levels $E_0$ (filled circle), $E_1$ (filled triangle), $E_2$ (open square), and $E_3$ (open inverted triangle)  
for the Al 100\% case in the Al/Au mixed site with $Z_{\rm Al}=xZ_{\rm Au}$ and $Z_{\rm Au}=0.01776159$.
(b) The CEF energy levels $\bar{E}_n$ $(n=0, 1, 2,$ and $3)$ averaged over all the configurations of the Al/Au mixed site. 
Notations are the same as (a). 
} 
\label{fig:E_x}
\end{figure}
%%%%%%%%%%%%%%%%%%%%%%%%%%%%%%%%%%%%%%%%%%%%%%%%

\begin{figure}[t]
\includegraphics[width=7cm]{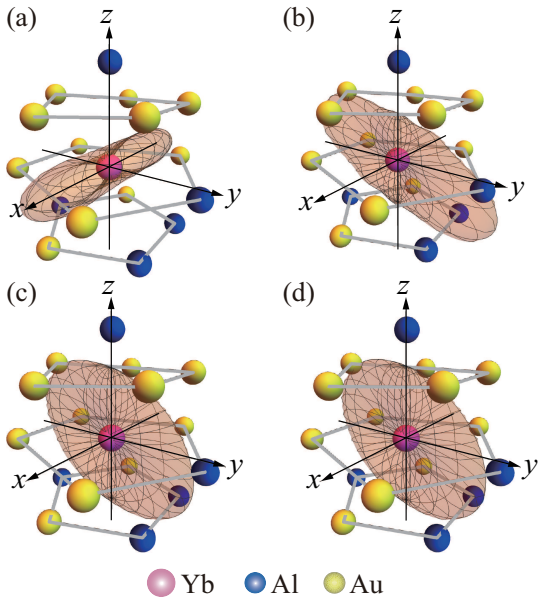}
\caption{(Color online) 
Square of the spherical part of the 4f wave function 
$|\psi_{0+}(\hat{\bm r})|^2$ of the CEF ground state (a) for $x=0.0$, (b) $x=1.0$, (c) $x=2.0$, 
and (d) $x=3.0$ for $Z_{\rm Al}=xZ_{\rm Au}$ and $Z_{\rm Au}=0.01776159$. 
} 
\label{fig:WF_x_3}
\end{figure}

\begin{figure}[t]
\includegraphics[width=8.5cm]{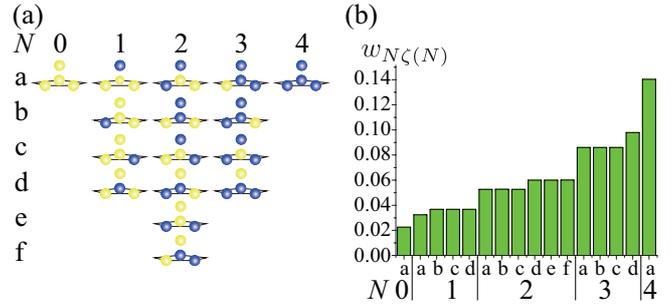}
\caption{(Color online) 
(a) Configurations of Al (blue) and Au (yellow) for the Al/Au mixed sites classified by the number of Al atoms. 
(b) Probability of each configuration of (a)~\cite{WM2018}. 
} 
\label{fig:AlAu}
\end{figure}

Figure~\ref{fig:E_x}(a) shows the $x$ dependence of the CEF energies $E_n$ $(n=0, 1, 2,$ and $3)$ with the Kramers degeneracy.
The energy levels close up around $x\approx 0.8$. 
To clarify how the CEF ground state evolves, we plot $|\psi_{n+}(\hat{\bm r})|^2$ in Fig.~\ref{fig:WF_x_3}.
Here, the spherical part of the 4f wave function is given by $\psi_{n\pm}(\hat{\bm r})\equiv\langle\hat{\bm r}|\psi_{n\pm}\rangle$ with $\hat{\bm r}$ being the unit vector for the radial direction. 
Since the Al  and Au ions are located symmetrically with respect to the mirror plane shown in Fig.~\ref{fig:Yb_local} as the $yz$ plane, the 4f wave function has the shape symmetric to the $yz$ plane. 
At $x=0.0$, $|\psi_{0+}(\hat{\bm r})|^2$ has a flat shape slightly tilted from the $xy$ plane as shown in Fig.~\ref{fig:WF_x_3}(a). 
As $x$ increases, the wave function is deformed as shown in Fig.~\ref{fig:WF_x_3}(b) for $x=1.0$.  
At $x=2.0$ and $3.0$, the wave function has the flat shape lying in the mirror plane i.e. $yz$ plane as shown in Figs.~\ref{fig:WF_x_3}(c) and \ref{fig:WF_x_3}(d).

%%%%%%%%%%%%  Fig.4  %%%%%%%%%%%%%%%%%%%%%%%%%%%%%%
%%%%%%%%%%%%%%%%%%%%%%%%%%%%%%%%%%%%%%%%%%%%%%%%

Note that the results shown in Figs.~\ref{fig:WF_x_3}(a)-\ref{fig:WF_x_3}(d) do not depend on the value of $Z_{\rm Au}$ itself. 
This is because the change in $Z_{\rm Au}$ merely multiply $H_{\rm CEF}$ a constant [see the 
%%%%%
%\textcolor{red}
{r.h.s.}
%%%%%
 of Eq.~(\ref{eq:Vcry})] so that the eigenstates do not change as far as $x$ is the same. 
This allows us to compare Fig.~\ref{fig:E_x} with the CEF energy levels observed by the inelastic neutron measurement. 
Namely, first $x$ is determined so as the measured energy levels to reproduce the ratios of energy differences $E_{n+1}-E_{n}$ for $n=0, 1, 2$  each other. Then, $Z_{\rm Au}$ is determined so as the absolute value of $E_n$ to reproduce the measured energy levels. 
In this way, the screened values of valences of Al and Au, i.e., $Z_{\rm Al}$ and $Z_{\rm Au}$ in metallic QC and AC can be determined.

%Experimentally, in the QC Yb$_{15}$Al$_{35}$Au$_{51}$, the energy difference $\Delta$ between the ground state and the first-excited state is estimated to be $\Delta\gsim 20$ ~K from the entropy obtained by integrating the electric specific heat by temperature as $S(T)=\int_0^TdT'C_{\rm e}(T')/T'$~\cite{Deguchi,D_PC}.  We note that $Z_{\rm Au}=0.01***$ gives $\Delta=20$~K. 

%\subsection{The effect of Al/Au mixed site}

Next, let us discuss the effect of the Al/Au mixed site. 
Considering the existence ratio of the Al/Au mixed sites in Fig.~\ref{fig:YbAlAu}~\cite{Ishimasa}, 
possible configurations of the Al/Au mixed site around Yb, which are illustrated in Fig.~\ref{fig:AlAu}(a), was calculated by the $10^{7}$ step of  random-number generation in ref.~\citen{WM2018}. 
Here, the Al/Au mixed site [i.e., the $i=2$, 4, 5, and 9th site in Fig.~\ref{fig:Yb_local}(a)] seen from the $y$ direction is illustrated in Fig.~\ref{fig:AlAu}(a).
The obtained probability of each configuration $w_{N\zeta(N)}$ is shown as the bar graph in Fig.~\ref{fig:AlAu}(b). 
Here, $N$ is the number of Al in the Al/Au mixed sites and $\zeta(N)$ specifies the configuration shown in Fig.~\ref{fig:AlAu}(a). 
%%%%%
%\textcolor{red}
{Although difference of the bond energy depending on each configuration was not explicitly taken into account in the calculation in ref.~\citen{WM2018}, the effect is considered to be reflected in Fig.~\ref{fig:AlAu}(b) resultantly since the existence ratios of the Al/Au mixed sites in the AC determined by the Rietveld structural analysis~\cite{Ishimasa} are reproduced by the configurations and weights shown in Figs.~\ref{fig:AlAu}(a) and \ref{fig:AlAu}(b) respectively.}
%%%%%

The CEF energy and wave function averaged over all the configurations of the Al/Au mixed site are obtained by 
\begin{eqnarray}
\bar{E}_n&=&\sum_{N=0}^{4}\sum_{\zeta(N)}w_{N\zeta(N)}E_{N\zeta(N),n}, 
\label{eq:Eav}
\\
\bar{\psi}_{n,\pm}(\hat{\bm r})&=&\sum_{N=0}^{4}\sum_{\zeta(N)}w_{N\zeta(N)}\psi_{N\zeta(N),n\pm}(\hat{\bm r}), 
\label{eq:WFav}
\end{eqnarray}
respectively, where 
$w_{N\zeta(N)}$ satisfies $\sum_{N=0}^{4}\sum_{\zeta(N)}w_{N\zeta(N)}=1$.
It is noted that 
as seen in Figs.~\ref{fig:AlAu}(a) and \ref{fig:AlAu}(b),  
each configuration not symmetric with respect to the mirror plane (i.e., $yz$ plane in Fig.~\ref{fig:Yb_local}) has corresponding configuration desymmetrized in the opposite direction with the same probability. 
Hence, after averaging the wave function in Eq.~(\ref{eq:WFav}), the shape of $|\bar{\psi}_{n,\pm}(\hat{\bm r})|^2$ becomes symmetric with respect to the mirror plane.   

%%%%%%%%%%%%  Fig.5  %%%%%%%%%%%%%%%%%%%%%%%%%%%%%%
%%%%%%%%%%%%%%%%%%%%%%%%%%%%%%%%%%%%%%%%%%%%%%%%

\begin{figure}[t]
\includegraphics[width=7cm]{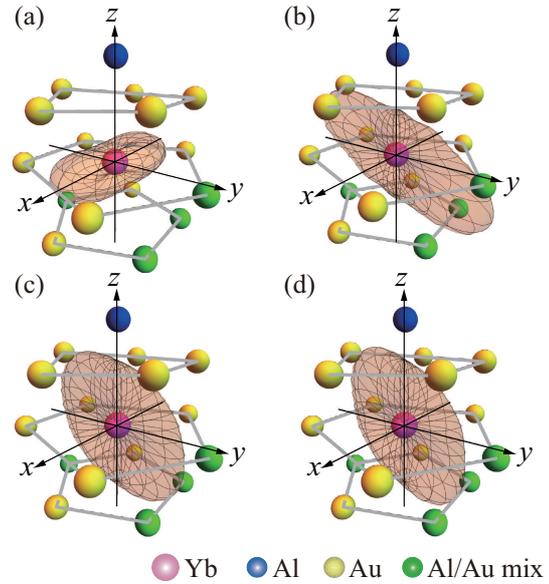}
\caption{(Color online) 
Square of the spherical part of the 4f wave function averaged over all Al/Au configurations shown in Fig.~\ref{fig:AlAu}(a)  
$|\bar{\psi}_{0+}(\hat{\bm r})|^2$ of the CEF ground state for (a) $x=0.0$, (b) $x=1.0$, (c) $x=2.0$, and (d) $x=3.0$ in $Z_{\rm Al}=xZ_{\rm Au}$. 
% and $Z_{\rm Au}=0.01776159$. 
} 
\label{fig:WFav}
\end{figure}

Figure~\ref{fig:E_x}(b) shows the $x$ dependence of the averaged CEF energy $\bar{E}_n$. The overall feature is similar to the Al 100\% case shown in Fig.~\ref{fig:E_x}(a). 
Figure~\ref{fig:WFav} shows square of the spherical part of the 4f wave function averaged over all Al/Au configurations shown in Fig.~\ref{fig:AlAu}(a) 
$|\bar{\psi}_{0+}(\hat{\bm r})|^2$ of the ground state for (a) $x=0.0$, (b) $x=1.0$, (c) $x=2.0$, and (d) $x=3.0$ in $Z_{\rm Al}=xZ_{\rm Au}$. 
% and $Z_{\rm Au}=0.01776159$. 
For $x\ge 1$, the shapes of the 4f wave function are almost similar to the Al 100\% case shown in Figs.~\ref{fig:WF_x_3}(b)-\ref{fig:WF_x_3}(d).  
For $x=0$, the 4f wave function has the almost flat shape lying in the $xy$ plane, as shown in Fig.~\ref{fig:WFav}(a).  
%while $|\psi_{0+}(\hat{\bm r})|^2$ for $x=0$ is slightly tilted from the $xy$ plane in the Al 100\% case shown in Fig.~\ref{fig:WF_x_3}(a).
%similarly to that for the Al 100\% case shown in Fig.~\ref{fig:WF_x_3}(a) but tilts in the opposite direction. 
%This is due to avoid the Al/Au mixed sites with positive charge of the Au ion causing the energy loss, which are taken into account in the average process. 
%For $x=3$, the averaged 4f wave function is lying in the mirror $(yz)$ plane as shown in Fig.~\ref{fig:WFav}(b). This result is almost similar to that for the Al 100\% case shown in Fig.~\ref{fig:WF_x_3}(d). 
Since the magnetic moment is generally aligned to the direction perpendicular to the flat shape of the 4f wave function, the results in Figs.~\ref{fig:WFav}(a)-\ref{fig:WFav}(d) indicate that the moment direction is changed from the nearly pseudo-5-fold-axis direction for $x=0$ to the perpendicular direction (i.e. $x$ axis) for $x=3$. Namely, the direction of the magnetic anisotropy changes according to $x$. It is noted here that the net magnetic moment at Yb is cancelled for the doubly degenerate CEF ground states $\bar{\psi}_{0\pm}(\hat{\bm r})$. 

%%%%%
%\textcolor{red}
{For $x=0$, the charges at the Al sites in the Al/Au mixed sites [i.e., the $i=2$, 4, 5, 9th site in Fig.2(a)] in Fig.4(a) are zero. By considering the configurations shown in Fig.~\ref{fig:AlAu}(a) in the averaging process of the Al/Au mixed sites, the charges distributed at the upper- and lower-pentagon Al/Au sites with respect to the $xy$ plane in Fig.~\ref{fig:WFav}(a) tend to be homogeneous. This makes the shape of the wave function flat lying in the $xy$ plane in comparison with that in Fig.~\ref{fig:WF_x_3}(a). On the other hand, for $x\gsim 1$, the charges at the Al sites in the Al/Au mixed sites in Figs.~\ref{fig:WF_x_3}(b)-\ref{fig:WF_x_3}(d) are positive. Hence, even by considering the configurations shown in Fig.~\ref{fig:AlAu}(a) in the averaging process of the Al/Au mixed sites, the charge $Z_{\rm Al}|e|$ is merely replaced with $Z_{\rm Au}|e|$, which resultantly gives no remarkable change in the shape of the wave function as shown in Figs.~\ref{fig:WFav}(b)-\ref{fig:WFav}(d). }
%%%%%

These results indicate that the analysis of the CEF assuming that the Al/Au mixed sites are all occupied by Al, which has the highest probability 
as shown in Fig.~\ref{fig:AlAu}(b), captures the qualitative feature especially in the $x\gsim 1$ regime but for details the averaging procedure by the Al/Au mixed configurations is necessary. 
Our study has clarified that the ratio of the valences of the screened ligand ions $x$ plays an important role in characterizing the CEF ground state.

%%%%%%%%%%%%  Fig.6  %%%%%%%%%%%%%%%%%%%%%%%%%%%%%%
%%%%%%%%%%%%%%%%%%%%%%%%%%%%%%%%%%%%%%%%%%%%%%%%

%\section{Discussion}

%In ref.~\citen{Hiroto}, from the inelastic neutron data combined with the analysis of the point charge model,  the valences of the screened Si and Au ions in the AC Au$_{70}$Si$_{17}$Tb$_{13}$ were estimated as $Z_{\rm Si}=0.578(3)$ and $Z_{\rm Au}=0.223(3)$ respectively. This gives the ratio $x=Z_{\rm Si}/Z_{\rm Au}\approx 2.59$. This approximately corresponds to the 4f wave function for $x=2\sim 3$ shown in Figs.~\ref{fig:WFav}(c) and \ref{fig:WFav}(d), which is lying in the mirror plane. Since the magnetic moment is generally aligned toward the direction perpendicular to the flat shape of the 4f wave function, the results shown in Figs.~\ref{fig:WFav}(c) and \ref{fig:WFav}(d) roughly correspond to the CEF ground state concluded in ref.~\citen{Hiroto}. The results in Figs.~\ref{fig:WFav}(b)-\ref{fig:WFav}(d) indicate that the magnetic moment is aligned to the direction perpendicular to the mirror plane for $1\lsim x\le 3$. On the other hand, the CEF ground state in refs.~\citen{Jazbec,Das} roughly corresponds to the result in the regime for $x< 0.8$ in Fig.~\ref{fig:E_x} [see Fig.~\ref{fig:WFav}(a)]. 
%Our study have clarified that the ratio of the screened ligand ions $x$ plays an important role for characterizing the CEF ground state. 

In general, the CEF is caused by the effects of the electro-static interaction and the hybridization. 
The present study has focused on the former effect. Now we briefly discuss the latter effect. 
It was reported that the quantum criticality disappears when Al is replaced with Ga in the QC Au$_{51}$Al$_{34}$Yb$_{15}$~\cite{Matsukawa2014}. 
This implies that the hybridization of 3p orbital at Al and 4f orbital at Yb contributes to the ground state. 
Actually, the theoretical analysis based on the extended periodic Anderson model in the AC showed that 
the hybridization of the 4f state at Yb and the 3p state at the nearest neighbor (N.N.) Al site  [i.e., the $i=9$th site in Fig.~\ref{fig:Yb_local}(a)] plays the important role in realizing the quantum critical state~\cite{WM2016,WM2018}. 
  
As shown in Fig.~\ref{fig:WFav}(d), the 4f wave function for $x=Z_{\rm Al}/Z_{\rm Au}=3$ extends to the N.N. Al site, which is expected to earn the lower energy when the hybridization effect is taken into account. 
Furthermore, this wave function also extends to the Al site along the pseudo 5-fold axis  [i.e., the $i=16$th site in Fig.~\ref{fig:Yb_local}(a)]. 
Since this Al site belongs to the 
%%%%%
%\textcolor{red}
{3rd} 
%%%%%
shell in Fig.~\ref{fig:YbAlAu}(d) with the 100\% existence probability, this state is also favorable when the hybridization effect is taken into account. 
%%%%%
%\textcolor{red}
{From these speculations, the features of the wave function which extends to the $i=9$ and 16th sites in Fig.~\ref{fig:Yb_local}(a) as in Figs.~\ref{fig:WFav}(c) and \ref{fig:WFav}(d) are expected to be relevant even after taking account of the effect of the hybridization as the CEF ground state. As for the CEF energy levels, the effect of the hybridization is estimated to give the $O((pf\sigma)^2/\Delta E_{\pm})$ contribution to the results in Fig.3 where $\Delta E_{+}$ $(\Delta E_{-})$ is the excitation energy of f$^1$ $\to$ f$^2$ (f$^1$ $\to$ f$^0$) in the hole picture.}
%%%%%
The analysis of the CEF by taking into account the hybridization effect is an interesting future subject.

%\section{Summary}

To summarize, 
on the basis of the point change model, we have formulated the CEF Hamiltonian $H_{\rm CEF}$ in the rare-earth based QC by the equivalent operators of the total angular momentum $J$.  
By diagonalizing $H_{\rm CEF}$ for $J=7/2$ in Au$_{51}$Al$_{34}$Yb$_{15}$, we have analyzed the CEF energies and the 4f wave functions. 
We have clarified that the ratio of valences of the screened ligand ions $x\equiv Z_{\rm Al}/Z_{\rm Au}$ plays the important role 
%%%%%
%\textcolor{red}
{in}
%%%%%
 characterizing the CEF ground state. The crossover of the CEF ground states occurs around $x\approx 0.8$. As $x$ decreases from $x=3$, the 4f wave function changes from the flat shape lying in the mirror plane to the almost flat shape perpendicular to the mirror plane at $x=0$.
Comparing our results with the future neutron measurement will make it possible to determine $x$ and $Z_{\rm Au}$, which identifies the CEF states. 
Notable is that $H_{\rm CEF}$ is formulated for general $J$ as in eq.~(\ref{eq:Vcry2}), which is applicable to the other rare-earth based QCs and ACs.
%%%%%
%\textcolor{red}
{It is interesting to analyze $H_{\rm CEF}$ in the Tb- and Tm-based ACs to be compared with experimental analysis of the CEF in refs.~\citen{Jazbec,Das,Hiroto}.  }
%%%%%

\begin{acknowledgment}
%\acknowledgment
%{\bf Acknowledgements} 

The authors thank T. Ishimasa for helpful discussions on the valences of Al and Au in the metallic AC.
This work was supported by JSPS KAKENHI Grant Numbers JP18K03542 and JP19H00648.
 
\end{acknowledgment}

\newpage
%%%%%%%%%%%%%%%%%%%%%%%%%%%%%%%%%%%%%%%%%%%%%%%%%%%%%%%%%%%%%%%%%%%%%%%%%%%%%%%%%%%%%%%%%%%%%%%

\onecolumn

\begin{center}
{\bf Supplemental Materials}
\end{center}

Stevens operators are expressed by operators of the total angular momentum $J_z$, $J_{+}\equiv (J_x+J_y)/2$, and $J_{-}\equiv (J_x-J_y)/(2i)$, as follows~\cite{Stevens,rotter}:  
\begin{eqnarray}
O_{2}^{2}(s)&=&\frac{-i}{2}(J_{+}^{2}-J_{-}^{2}),\\
O_{2}^{1}(s)&=&\frac{-i}{4}[J_{z}(J_{+}-J_{-})+(J_{+}-J_{-})J_{z}],\\
O_{2}^{0}&=&3J_{z}^{2}-J(J+1),\\
O_{2}^{1}&=&\frac{1}{4}[J_{z}(J_{+}+J_{-})+(J_{+}+J_{-})J_{z}],\\
O_{2}^{2}&=&\frac{1}{2}(J_{+}^{2}+J_{-}^{2}),\\
O_{4}^{4}(s)&=&\frac{-i}{2}(J_{+}^{4}-J_{-}^{4}),\\
O_{4}^{3}(s)&=&\frac{-i}{4}[(J_{+}^{3}-J_{-}^{3})J_{z}+J_{z}(J_{+}^{3}-J_{-}^{3})],\\
O_{4}^{2}(s)&=&\frac{-i}{4}[(J_{+}^{2}-J_{-}^{2})(7J_{z}^{2}-J(J+1)-5)+(7J_{z}^{2}-J(J+1)-5)(J_{+}^{2}-J_{-}^{2})],\\
O_{4}^{1}(s)&=&\frac{-i}{4}[(J_{+}-J_{-})(7J_{z}^{3}-(3J(J+1)+1)J_{z})+(7J_{z}^{3}-(3J(J+1)+1)J_{z})(J_{+}-J_{-})],\\
O_{4}^{0}&=&35J_{z}^{4}-(30J(J+1)-25)J_{z}^{2}+3J^{2}(J+1)^{2}-6J(J+1),\\
O_{4}^{1}&=&\frac{1}{4}[(J_{+}+J_{-})(7J_{z}^{3}-(3J(J+1)+1)J_{z})+(7J_{z}^{3}-(3J(J+1)+1)J_{z})(J_{+}+J_{-})],\\
O_{4}^{2}&=&\frac{1}{4}[(J_{+}^{2}+J_{-}^{2})(7J_{z}^{2}-J(J+1)-5)+(7J_{z}^{2}-J(J+1)-5)(J_{+}^{2}+J_{-}^{2})],\\
O_{4}^{3}&=&\frac{1}{4}[(J_{+}^{3}+J_{-}^{3})J_{z}+J_{z}(J_{+}^{3}+J_{-}^{3})],\\
O_{4}^{4}&=&\frac{1}{2}(J_{+}^{4}+J_{-}^{4}),\\
O_{6}^{6}(s)&=&\frac{-i}{2}(J_{+}^{6}-J_{-}^{6}),\\
O_{6}^{5}(s)&=&\frac{-i}{4}[(J_{+}^{5}-J_{-}^{5})J_{z}+J_{z}(J_{+}^{5}-J_{-}^{5})],\\
O_{6}^{4}(s)&=&\frac{-i}{4}[(J_{+}^{4}-J_{-}^{4})(11J_{z}^{2}-J(J+1)-38)+(11J_{z}^{2}-J(J+1)-38)(J_{+}^{4}-J_{-}^{4})],\\
O_{6}^{3}(s)&=&\frac{-i}{4}[(J_{+}^{3}-J_{z}^{3})(11J_{z}^{3}-(3J(J+1)+59)J_{z})
\nonumber
\\
&+&
(11J_{z}^{3}-(3J(J+1)+59)J_{z})(J_{+}^{3}-J_{z}^{3})],\\
O_{6}^{2}(s)&=&\frac{-i}{4}[(J_{+}^{2}-J_{-}^{2})(33J_{z}^{4}-(18J(J+1)+123)J_{z}^{2}+J^{2}(J+1)^{2}+10J(J+1)+102)
\nonumber 
\\
&+& 
(33J_{z}^{4}-(18J(J+1)+123)J_{z}^{2}+J^{2}(J+1)^{2}+10J(J+1)+102)(J_{+}^{2}-J_{-}^{2})],\\
O_{6}^{1}(s)&=&\frac{-i}{4}[(J_{+}-J_{-})(33J_{z}^{5}-(30J(J+1)-15)J_{z}^{3}+(5J^{2}(J+1)^{2}-10J(J+1)+12)J_{z})
\nonumber
\\
&+&
(33J_{z}^{5}-(30J(J+1)-15)J_{z}^{3}+(5J^{2}(J+1)^{2}-10J(J+1)+12)J_{z})(J_{+}-J_{-})],\\
O_{6}^{0}&=&231J_{z}^{6}-(315J(J+1)-735)J_{z}^{4}+(105J^{2}(J+1)^{2}-525J(J+1)
\nonumber
\\
&+&
294J_{z}^{2}-5J^{3}(J+1)^{3}+40J^{2}(J+1)^{2}-60J(J+1),\\
O_{6}^{1}&=&\frac{1}{4}[(J_{+}+J_{-})(33J_{z}^{5}-(30J(J+1)-15)J_{z}^{3}+(5J^{2}(J+1)^{2}-10J(J+1)+12)J_{z})
\nonumber
\\
&+&
(33J_{z}^{5}-(30J(J+1)-15)J_{z}^{3}+(5J^{2}(J+1)^{2}-10J(J+1)+12)J_{z})(J_{+}+J_{-})],\\
O_{6}^{2}&=&\frac{1}{4}[(J_{+}^{2}+J_{-}^{2})(33J_{z}^{4}-(18J(J+1)+123)J_{z}^{2}+J^{2}(J+1)^{2}+10J(J+1)+102)
\nonumber
\\
& &+(33J_{z}^{4}-(18J(J+1)+123)J_{z}^{2}+J^{2}(J+1)^{2}+10J(J+1)+102)(J_{+}^{2}+J_{-}^{2})],
\\
O_{6}^{3}&=&\frac{1}{4}[(J_{+}^{3}+J_{z}^{3})(11J_{z}^{3}-(3J(J+1)+59)J_{z})
\nonumber
\\
&+&
(11J_{z}^{3}-(3J(J+1)+59)J_{z})(J_{+}^{3}+J_{z}^{3})],\\
O_{6}^{4}&=&\frac{1}{4}[(J_{+}^{4}+J_{-}^{4})(11J_{z}^{2}-J(J+1)-38)+(11J_{z}^{2}-J(J+1)-38)(J_{+}^{4}+J_{-}^{4})],\\
O_{6}^{5}&=&\frac{1}{4}[(J_{+}^{5}+J_{-}^{5})J_{z}+J_{z}(J_{+}^{5}+J_{-}^{5})],\\
O_{6}^{6}&=&\frac{1}{2}(J_{+}^{6}+J_{-}^{6}). 
\end{eqnarray}
\end{document}